# Security, Privacy and Safety Risk Assessment for Virtual Reality Learning Environment Applications

Aniket Gulhane[a], Akhil Vyas[b], Reshmi Mitra[a], Roland Oruche[a], Gabriela Hoefer[c],
Samaikya Valluripally[a], Prasad Calyam[a], Khaza Anuarul Hoque[a]
University of Missouri-Columbia[a], Missouri University of Science and Technology[b], Samford University[c],
{arggm8, rro2q2, svbqb}@mail.missouri.edu, avgm3@mst.edu, ghoefer@samford.edu, {mitrare, calyamp, hoquek}@missouri.edu

*Abstract*—Social Virtual Reality based Learning Environments (VRLEs) such as vSocial render instructional content in a three-dimensional immersive computer experience for training youth with learning impediments. There are limited prior works that explored attack vulnerability in VR technology, and hence there is a need for systematic frameworks to quantify risks corresponding to security, privacy, and safety (SPS) threats. The SPS threats can adversely impact the educational user experience and hinder delivery of VRLE content. In this paper, we propose a novel risk assessment framework that utilizes attack trees to calculate a risk score for varied VRLE threats with rate and duration of threats as inputs. We compare the impact of a well-constructed attack tree with an adhoc attack tree to study the trade-offs between overheads in managing attack trees, and the cost of risk mitigation when vulnerabilities are identified. We use a vSocial VRLE testbed in a case study to showcase the effectiveness of our framework and demonstrate how a suitable attack tree formalism can result in a more safer, privacy-preserving and secure VRLE system.

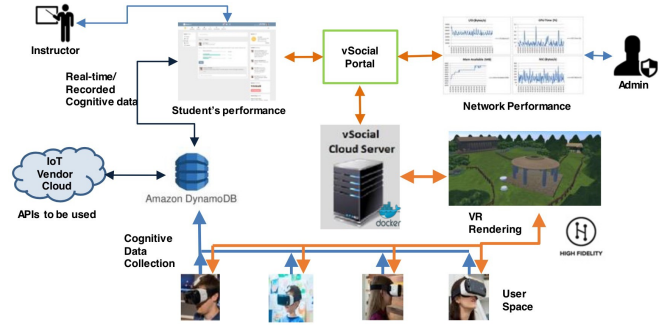

Fig. 1: vSocial system showing the cloud server used for real-time student learning session management.

## I. INTRODUCTION

Ensuring human safety along with relevant security and privacy mechanisms are major challenges in emerging Virtual Reality based Learning Environments (VRLEs) such as vSocial [1] for youth with learning disabilities. In order to assess student engagement levels, VRLEs use emotion tracking sensors whose data along with student learning progress are stored in a cloud platform. In spite of their promising potential as immersive instructional platforms, the real-time delivery of learning material poses security and privacy threats. This is a multi-modal system as shown in Fig. 1, which is built upon rendering 3-dimensional visualizations based on dynamic human computer interactions. All these components increase the attack surface area and negligence to address threats may result in negative impacts such as alteration of instructional content, compromise of learning outcomes or granting users unnecessary "false strike" penalties. This in turn may result in poor student engagement in ongoing classroom sessions.

Studies such as [2] highlight the importance of security and privacy issues in VR devices. However, there are few efforts in evaluating the threat scenarios in such dynamic, complex and large-scale collaborative systems, thereby hindering the design of secured and usable VRLEs. To the best of our knowledge, our work is one of the first systematic studies to evaluate the security and privacy concerns along with the quality of human experience in VRLEs. The focus of our work is to categorize threats based on three orthogonal aspects in VR applications: security, privacy and safety (SPS). We define *security* as the robustness of the VR system against various attacks, *privacy* as the protection and secrecy over data sensitivity, and *safety* as the disruption in the system that compromises the user's overall well-being.

In this paper, we address above issues by proposing a "risk assessment framework" utilizing attack trees to formalize and evaluate internal and external vulnerabilities, and prioritize threats based on impact. Attack trees have been used extensively in threat modeling, and for understanding intrusion/attack scenarios by their quantitative analysis. We apply the concept of attack trees to VRLEs to study how particular events can functionally harm specific SPS factors using a hierarchical visual representation of potential vulnerabilities and threats. With the outputs of attack trees, we quantify the risk score for each SPS factor using frequency rates and duration of threats as parameters.

We use the vSocial [1] system, a VRLE for youth with Autism Spectrum Disorder (ASD), as a case study to demonstrate the effectiveness of our framework. We introduce three

This material is based upon work supported by the National Science Foundation under Award Numbers: CNS-1647213 and CNS-1659134. Any opinions, findings, and conclusions or recommendations expressed in this publication are those of the authors and do not necessarily reflect the views of the National Science Foundation.

sample attack scenarios that cover various SPS aspects such as malicious network discrepancies, packet loss, and packet sniffing to measure the impact of SPS threats on VRLE applications. We use the measures from these simulations as inputs for attack tree nodes to demonstrate the difference in efficiency of a complete/well-constructed and incomplete/adhoc attack tree. Corresponding risk score analyses provide insights about system component vulnerabilities, which influence the VRLE policy management for threat mitigation.

The main contributions of this paper can be summarized as follows:
*(1)* Risk assessment with a comprehensive threat model for VRLEs that includes threats to system robustness, user information and well-being.
*(2)* SPS attack trees to highlight the inter-relationship of threats and the corresponding system component(s), which ultimately affects the risk score.
*(3)* Trade-off analysis between attack tree qualities to obtain insights for component vulnerability to inform VRLE design.

The remainder of the paper is organized as follows: Section II discusses related works on VRLE, as well as threats to security, privacy and safety in Internet-of-Things (IoT) devices. Section III formally states the SPS problem using threat formalization and risk assessment. Section IV presents performance evaluation results of our risk assessment framework. Section V concludes the paper.

## II. RELATED WORK

### A. Virtual Reality Learning Environments

Prior works on VRLEs have been developed for immersive and special education purposes. In vSocial [1], a Social Competence Intervention (SCI) curriculum [3] is delivered via a system that consists of various modules such as: VR rendering, web applications and classroom portal with instructional content hosted as web pages. The vSocial server provides functionalities such as access control for users, session management, network performance tracking, session progress, and user management, making it a critical target in the system.

A recent study [4] on security and privacy challenges in Augmented Reality (AR) and VR discusses the threat surface area for educational initiatives without characterizing the attack impact, however, there are very few scholarly works on the interplay of SPS factors from a usability perspective (i.e., learning experience in our use case). In VRLE, the instructor plays a critical role by changing the pace of curriculum delivery and educational VR content in the virtual classroom based on continuous student evaluation (e.g., tokens, passes or strikes). The VRLE administrator also occupies an elevated position as he or she controls the user data of several sessions. Thus, any disruption due to masquerading or spoofing by attacker with malicious intents [5] on the instructor's VR content or administrator privileges will compromise the learning activities in such a collaborative (virtual) space.

The geographically distributed students setup also makes the system susceptible to Distributed Denial of Service (DDoS) attacks as it compromises with the availability of the learning environment, impacts real-time data collection and student performance visualization. Even intermittent network discrepancy attacks can cause disruptions during sessions leading to cybersickness and unnecessary "false strikes." Estimating the impact of these threats is complex and challenging in dynamic VRLEs because of the multi-modal distributed system and high volume of real-time data. Our work proposes a novel approach in addressing and evaluating the SPS threats with a novel risk assessment framework in context of exemplar VRLEs such as vSocial in a case study.

### B. Security, Privacy and Safety

We present a summary of SPS issues in emerging technologies such as IoT as discussed in [6], which in particular, comprises of an exhaustive compilation of potential security and privacy threats and challenges. Another comprehensive study about this topic is presented in [7]. A typical VRLE application is susceptible to similar vulnerabilities along with human well-being, which has not been previously addressed in existing works. It is important to note that we are not exploring the trustworthiness, but rather focusing on the effect of VR usability due to potential security and privacy attacks.

Authors in [8] discuss challenges in security and privacy for Augmented Reality (AR) applications and explore opportunities for securing AR systems without much discussion about the safety aspects. Although threats in AR and other IoT systems are also relevant in VR applications, VR threats differ from AR threats because of the complete user immersion in a virtual world. Multiple attacker models in [9] for threats in security and privacy for distributed IoT systems focus on threats such as DoS, physical damage, eavesdropping, etc. They suggest countermeasures but without evaluation studies. A survey in [10] classifies the security and privacy attacks in IoT systems, and discusses security issues in different layers i.e., in application, network, transport and perception.

Previous research has examined privacy issues in AR, fog and mobile computing. The AR browser in [11] examines in depth about the vulnerabilities and requirements for mobile devices without any significant evaluation of proposed approaches. Works about privacy attacks for fog computing [12] discuss issues such as trust and authentication, data storage, location and usage privacy. A thorough survey on mobile computing privacy threats [13] highlights the trade-off between functionality and privacy. Some major threats which are also applicable to our VRLE include: lack of transparency, tracking, leaks from (mobile) sensor, among others.

A seminal work on safety issues for virtual environments [14] established that human performance efficiency is affected by task and user characteristics. Existing works such as [15], compared a virtual environment in a display monitor with Head-Mounted Display (HMD) to establish its correlation with cybersickness. Most recently, works such as [16] highlight the problems about overexposure and cybersickness in VRLEs for training youth with ASD. Considering the existing SPS research in such VRLE applications, our work not only considers security and privacy, but also safety threats in a single comprehensive risk assessment framework.

*C. Risk Assessment Approaches*

Risk assessment is performed by analyzing threat parameters (influence on asset, recovery cost, probability of occurrence) to determine its risk priority. This can help in mitigating the risks and the design of defense mechanisms. Earlier works on attacks in cyber-physical systems in [17] perform a risk assessment for supervisory control and data acquisition (SCADA) and distributed control systems (DCSs) by quantitatively determining probability and impact of attack. This helps them perform risk reduction by designing targeted countermeasures. Attack tree is a hierarchical model about threats and respective attack scenarios. Risk assessment using well-constructed attack trees can prove to be a cost-effective approach in designing a protected system.

Authors in [18] explain the theory and importance of attack trees. In a practical use case of online banking [19], the authors show how to use attack (and protection) trees to explain vulnerabilities, and develop a protection mechanism for system security. A work on smart cars using attack tree analysis [20] proposes a risk assessment framework to efficiently formulate the security measures. For analyzing security threats in ATMs [21], the authors construct attack-defense trees. Our work builds upon their formalization of attack trees in the context of SPS factors in a collaborative VRLE with realistic application test scenarios and VR content.

## III. THREAT FORMALIZATION AND RISK ASSESSMENT

To facilitate collaboration among students distributed over multiple locations, social VRLE applications such as vSocial [1] demand continuous and secured interoperability with entities (e.g., network-connected edge cloud) that requires large-scale capture, processing, and visualization of sensor data streams. These inherent challenges demand novel techniques for threat modeling of SPS factors in such complex multi-modal systems. In our proposed work, we provide threat formalization and risk assessment of the virtual reality learning environments using vSocial as a case study. We consider the potential threats pertaining to the vSocial server as the critical attack target, along with the respective SPS factors as shown in Fig. 2. We use these threat-to-system component mappings to design our attack trees, which ultimately quantify the risk score for each of the VRLE application system modules.

*A. Threat Model*

A threat model is defined as a framework which details internal and external vulnerabilities, as well as objectives and countermeasures [22]. Threat models are utilized across various disciplines such as cloud computing [23], health records [24], and storage [25]. Threat model for a multi-modal system such as VRLE can provide a systematic analysis of possible threats and help identify any module that is highly vulnerable to attacks.

Rather than exploring threats in the entire vSocial system, we consider the vSocial server as a critical target (i.e., the trusted computing base) as it executes several functionalities such as: rendering controls, visualization, web applications, storage, as shown in Fig. 1. The session permissions refer to users' ability to access session resources. The storage deals with transit data, which is any data collected in real-time such as emotion data or network performance measurements, and static data, which refers to user data and progress reports. The visualization functionality allows for display of real-time data so that the instructor or administrator can view it on a web portal. The rendering controls enable the instructor to invite students (other users) to join the VR class. A compromise in the security of these modules can leak confidential information or could compromise the integrity of the entire VRLE system. In the following paragraphs, various SPS attacks on server components are summarized:

*1) Security:* All VRLE systems are open to security attacks that can compromise integrity and performance. Session failure results from malicious activities carried out by attacker to crash VR sessions such as Denial of Service (DoS) attacks [9]. Threats such as Elevation of Privilege (EOP) provide attackers elevated access to sessions and activities that enable them to modify system contents or add malicious files. A typical VRLE system collects large amounts of sensitive data, which are susceptible to manipulation through data tampering performed via unauthorized channels. Furthermore, data integrity can be compromised by insertion of malicious code to modify session entities or data, or even change system configuration or access policies. Also, network attacks, such as DoS or DDoS result in system crashes and data unavailability. Moreover, impersonation attacks can occur when impostors login with stolen credentials, and access sensitive user information.

*2) Privacy:* Threats to privacy impact VRLE data confidentiality. Attacks such as eavesdropping allow an attacker to access confidential information via packet sniffing. Also through shoulder surfing, attacker can gain access to user authorization information through screen or hand movement observation in VR sessions. Furthermore, data security breaches including tampering with static and transit data allows the attacker to gain access to user credentials and real-time data. Informed consent is also a concern if users are not notified about what data is being collected from them. Improper disposal of data can also compromise privacy due to deleted information still residing in the server, putting the users confidentiality at risk.

*3) Safety:* We consider safety threats to be factors that directly impact user well-being. For instance, session takeover allows an attacker to control the VR rendering, impacting user activity in a VR session. Moreover, any network discrepancy initiated by an attacker can cause sudden changes in VR rendering leading to user disorientation. Unintentional activities can also cause safety issues. For example, computer bugs can cause glitches within the VR rendering software, resulting in sudden differences in visuals inside the headset. Extended sessions can cause cybersickness [15] due to an individual being forced to stay in a VR session for an extended period of time.

*B. Risk Assessment*

We designed a risk assessment framework to examine consequences of undesirable events, predict the likelihood of an attack and prioritize the threats accordingly. In subsequent paragraphs we formalize our risk assessment framework using attack tree formalisms for threat scenarios described above.

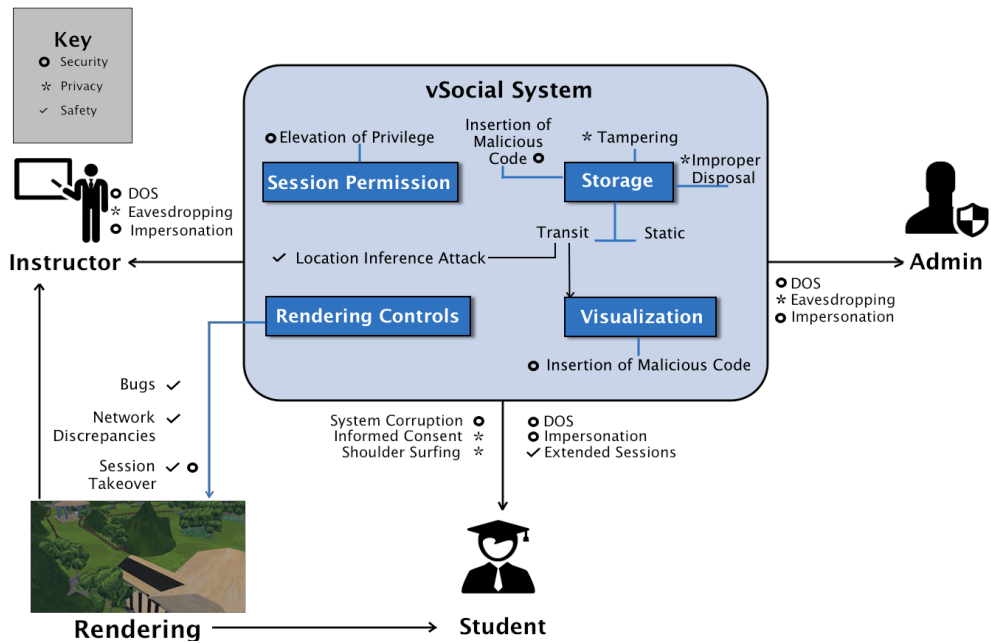

Fig. 2: Threat model representing formalization and classification of Security, Privacy and Safety threats originating in the vSocial VRLE server hosting the virtual reality content.

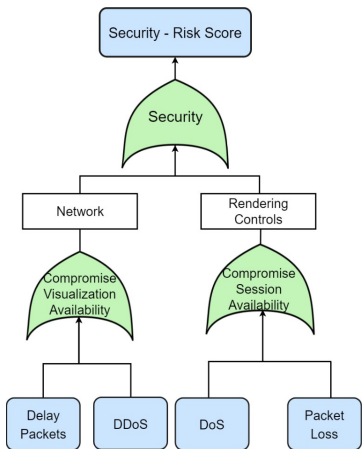

Fig. 3: A sample security attack tree showing denial of service attacks on network and rendering controls.

*Attack Trees*: They provide a formal, hierarchical structure to represent the relationship among possible system vulnerabilities and their respective attack scenarios. The format of these diagrams is the tree structure with 'target as the root node' and the 'leaf nodes describe the different activities carried out by the attacker. For the parent node scenario to be true, it is important that the child node logical condition (AND/OR) is also true, which is an elegant way to abstract multi-level threat impact. An attack tree considers the probabilistic effect of a particular event when it is influenced by a combination of multiple events. In our work, we apply this concept to threats on the vSocial server for analyzing the *risk score* based on the probability of occurrence of any threat event.

There are several tools to develop attack trees such as ADTooL, SecurITree [26], among others which formalize attack trees as part of risk assessment. SecurITree specifically shows its flexibility by considering several input parameters such as: rates, weights, and counter measure nodes at every level of the tree. For our proposed risk, we utilize frequency rate and duration of attacks. Frequency rates refer to the number of times an attack occurs in a specific time period, and the duration is the timespan of the attack on the VRLE system. Other functionalities such as counter measure node could also be considered within a VRLE application setup context.

*Risk Assessment for vSocial*: Our attack tree for risk assessment covers direct and indirect threats that manifest on the vSocial server and maps them back to system component that was the point of attack origin. The sample tree as shown in Fig. 3 is an example for DoS attacks impacting two separate system components - network and VR rendering controls. Using this sample tree, we can demonstrate the attack impact on system security. Propagating up in the tree, either DoS *or* delayed packets can disrupt network security, whereas any DoS or packet loss on session availability can cause major issues in VR rendering controls. The overall impact is a security compromise and hence marked as the root node. With the same logic, sample attack trees are shown for all the SPS factors - security in Fig. 4, privacy in Fig. 5 and safety in Fig. 6, respectively. The security tree explains how attacks on VR space, storage, network, VR rendering controls and visualization can impact system robustness. The privacy tree exemplifies that loss of user information can happen at either storage, VR space or network. The safety tree represents that user well-being can be compromised due to issues in VR space, VR rendering, location information, network or extended sessions.

The output of the attack tree is the "probability of occurrence," which is a popular risk metric. Additional metrics include distribution of loss or cost of attacks, however they are

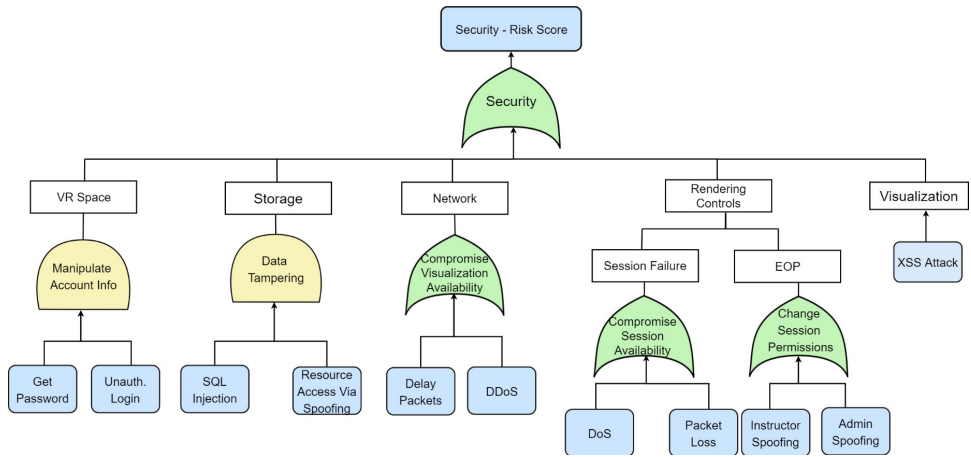

Fig. 4: Security attack tree showing threats in VR space, storage, network, rendering controls and visualization.

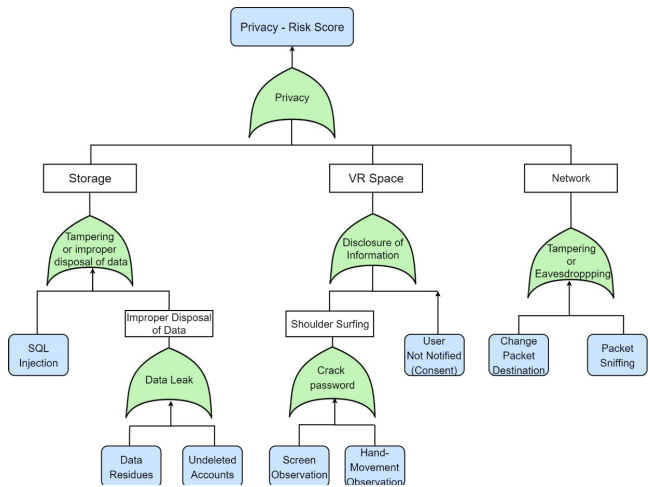

Fig. 5: Privacy attack tree showing threats in storage, VR space and network.

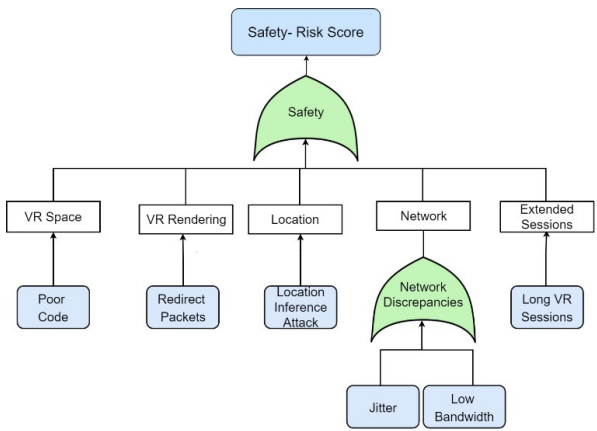

Fig. 6: Safety attack tree showing threats on user safety concerns.

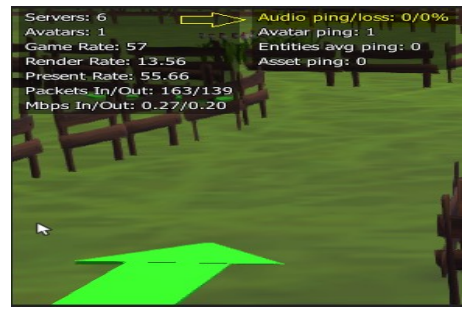

(a) Normal virtual reality rendering with no packet loss

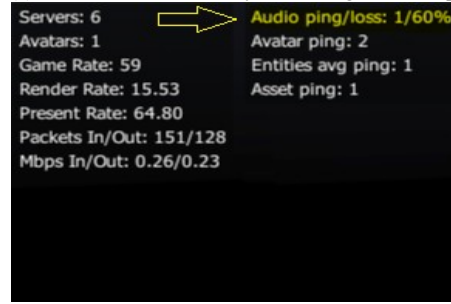

(b) Virtual reality rendering is completely disrupted due to packet loss

Fig. 7: Effect of packet loss on virtual reality rendering.

difficult to measure within a pre-determined time period. The probability of occurrence combined with the individual threat impact helps us perform the risk assessment by generating the *risk score* for the SPS threats. In other words, this will quantify the risk score associated with each of the modules in the vSocial server as shown in Figs. 4, 5 and 6. The higher the SPS risk score for a particular threat, higher is the corresponding risk associated with it. Based on these values, the VRLE administrator can get better visibility about system component vulnerability for a variety of threats and subsequently plan defense mitigation strategies for a safer, privacy-preserving and a more secure VRLE system.

## IV. Performance Evaluation

In this section, we first outline our data collection experiments that provide insights for our risk score evaluations. Following this, we study the importance of system design with complete/adhoc quality attack trees towards understanding the threat vulnerabilities in a well-defended VRLE system.

### A. System Testbed and Measurement Tools

The vSocial testbed described in [1] is a cloud-based and high-speed network-enabled application hosted on the Global Environment for Network Innovations (GENI) cloud rack [27], with the VR content developed in High Fidelity [28]. The users connect to the virtual classroom via HMD devices such as HTC Vive and Oculus Rift. The measurement tools used for threat simulations in our testbed are summarized below:

(1) *Steam* - Online game platform [29] which also contains an inbuilt VR system called steam VR. We utilize its frame timing tool to record frame rates in real-time under different network conditions.

(2) *Netlimiter* - Internet traffic control and monitoring tool for network monitoring and tuning in application under study. We use it to simulate DoS attacks that create network discrepancies on our High Fidelity based VRLE application, and measure the corresponding frame rate and other impacts on VR content rendering.

(3) *Wireshark* - A free and open source network protocol analyzer used for packet analysis, network troubleshooting and monitoring. We use it to capture packets being sent to-and-from our High Fidelity based VRLE server to demonstrate possible data leaks resulting from the capturing of application packets.

(4) *Clumsy 0.2* - A Windows based tool [30] that allows controlling network conditions including features such as lag, drop, throttle, or tamper of live packets. We specifically utilize the drop feature to drop a specific percentage of our packets to see the effect on our VRLE application performance.

### B. Test Case Evaluation

Our experimental attack scenarios include network discrepancy and packet loss for security, packet sniffing for privacy and their impacts on safety. Their real-time measurements are provided as inputs in our attack tree to generate a realistic *risk score*. For our work, the duration measurements from packet loss were supplied as inputs to the security attack tree as shown in Fig. 4.

*1) Security:* Using Clumsy 0.2, we changed the percentage of packets dropped to measure the time for a complete crash of a VRLE environment and recorded its effect on the VRLE environment as shown in Table I. The corresponding VR world screens with and without the presence of severe packet loss are shown in Fig. 7a and Fig. 7b, respectively.

For network discrepancy, we considered different bandwidth qualities - high (normal), medium and low (see Table II) for upload and download speed to observe the frame rate and High Fidelity content download time. The results summarized in Table II show that any upload speed below 30 Kbps resulted in High Fidelity crashing. We observed similar impacts on VR headset content rendering. Fig. 8a shows the frame rates inside the headset under high network performance and showcases a smooth VR rendering. Fig. 8b displays the same CPU performance under a medium network condition, where the graph now contains a red coloring as well. This means that the frame rate is not really being affected, but that the High Fidelity VRLE application is using more than its allotted CPU budget. The GPU performance shows similar trends, but has not been shown for space constraints. The most drastic case in Fig. 8c displays the frame rate under a low bandwidth with disruptive fluctuations that makes the experience inside the headset undesirably rough and jittery.

TABLE I: Impact of packet loss on virtual reality learning environment with metrics including packet drop percentage and average crash time.

| Packet Drop | Crash Time (secs) | Impact |
| --- | --- | --- |
| 20% | N/A | Slow content Rendering |
| 40% | 123 | Crash |
| 60% | 36 | Crash |
| 80% | 12 | Crash |

TABLE II: Upload and Download Speeds in Kbps for different bandwidth qualities.

|  | High | Medium | Low |
| --- | --- | --- | --- |
| Screenshots | Fig. 10a | Fig. 10b | Fig. 10c |
| Download Time (secs) | 12.2 | 33.3 | >300 |
| Upload Speed (Kbps) | 69 | 40.04 | <30 |
| Download Speed (Kbps) | 134 | 86 | 27.66 |

*2) Privacy:* In order to simulate packet sniffing attacks, we captured a subset of packets being sent to/from our High Fidelity VRLE server IP address using Wireshark. From the stream for captured packets, we viewed the avatar information and confidential host as well as server details as shown in Fig. 9. This demonstrated that any packet containing confidential information about the user or application can be captured and deciphered. This becomes a serious risk compromising privacy, especially without a secured network protocol.

*3) Safety:* For the simulation of safety attacks, we conducted a usability study introducing threats such as session failure and network discrepancy. Through this study, we measured the user Quality of Experience (QoE) [31]. It is important to note that any security and privacy threats can also have safety consequences. For example, reducing the bandwidth can cause sudden changes in the VR content, which could severely impact the users' educational experience. A normal environment such as Fig. 10a can suddenly lose content as shown in Fig. 10b and Fig. 10c. This in turn could highly disorient users and increase confusion and frustration levels. For our usability study, five participants entered the VRLE and their experience was measured under two conditions. They represented the two scenarios of control case with no threats and security breach.

Users were given a post simulation questionnaire, where they answered 28 questions on a popular psychometric 7-point Likert scale (strongly agree to strongly disagree), and seven additional open-ended questions further examining their

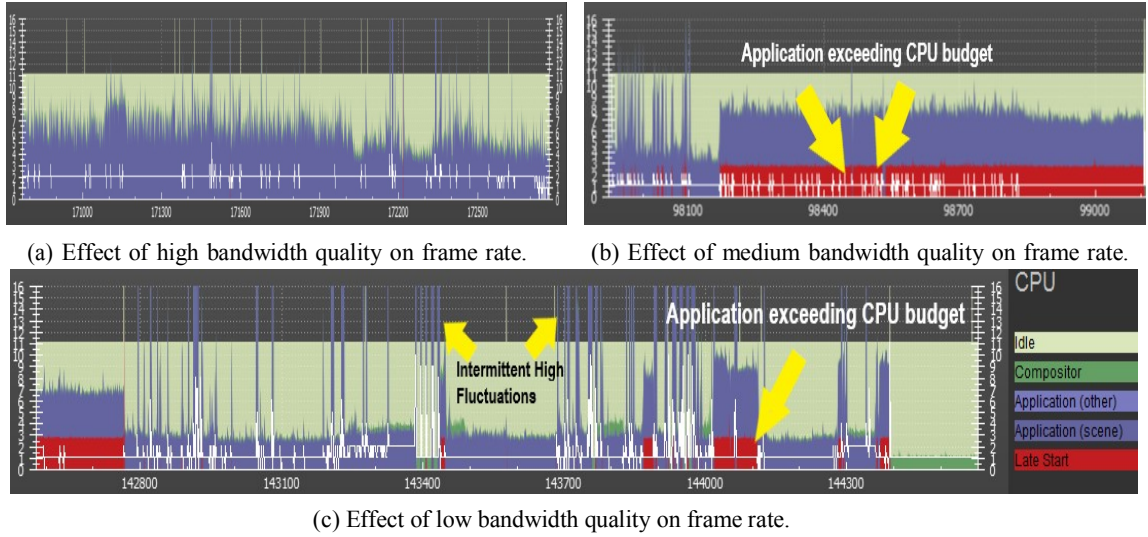

(a) Effect of high bandwidth quality on frame rate.

(b) Effect of medium bandwidth quality on frame rate.

(c) Effect of low bandwidth quality on frame rate.

Fig. 8: Effect of high, medium and low bandwidth quality on frame rate using Frame Timing tool in SteamVR; Effects of these conditions on virtual reality rendering are shown in Fig. 10.

Fig. 9: Packet Sniffing using Wireshark showing avatar and host server information getting disclosed.

experience. We analyzed the statistically significant differences between the two conditions (breach and no breach) and displayed the results in Fig. 11 to demonstrate a clear difference specifically in dizziness, confusion, control, and attitude towards the VRLE.

This further proves the benefits of conducting risk assessments by using attack trees and through characterization of user experience under security and privacy threat conditions.

### C. Attack Tree Evaluation

Based on the simulation for packet loss as shown in Table I, we used time measurements as duration input values and assumed frequencies to generate probabilities of occurrence for the security attack tree shown in Fig. 4. We also added some intermediary values and obtained the results in Fig. 12, and observed a directly proportional relation between duration and frequency of attack with the risk score.

We next present the characteristics required for the formulation of an effective, complete attack tree. For designing a complete attack tree and getting a realistic risk score, it is imperative to also consider indirect effects of threats and all system components. An adhoc attack tree alternately underrepresents the system vulnerabilities thus having a reduced risk score. We reiterate our earlier observation that lower risk score represents low susceptibility to threats, and hence requires less sophisticated defense mechanisms. This seems to be operationally simple and also cost effective for a VRLE administrator. An adhoc attack tree quality, however, fails to expose the comprehensive attack scenarios giving a false impression of a well-defended system or a system that requires sophisticated cost-prohibitive defense mechanism.

TABLE III: Impact scale of threats on vSocial safety.

| Attack Events | Scale of Impact |
|---|---|
| Redirect Packets to Malicious Server | 4 |
| Poorly Written Code | 1 |
| Location Inference attack | 2 |
| Jitter | 3 |
| Low Bandwidth | 4 |
| Long VR sessions | 3 |

For example in vSocial, if someone gains unauthorized access to the VRLE administrator or instructor accounts, other threats such as data tampering or elevation of privilege (EOP) can be triggered as shown in Fig. 4. If indirect implications of some threats are not considered in the formalization of the attack tree, the complete picture of possible threats will not be captured as illustrated in Fig. 3. In the second example, if all the threats in storage module are not addressed, then the module becomes vulnerable to unaccounted attacks. This could result in disclosing confidential user data or session information.

For the characteristic evaluation, attack trees are tested for different duration of attacks ($d_1$ = 20 min, $d_2$ = 10 min and $d_3$ = 5 min), and frequency of attacks ($f_1$ = 5/day, $f_2$ = 3/day and $f_3$ = 2/day). Fig. 13 verifies that the SPS risk score increases as more threats are addressed.

The main participants in VRLE system are the students, instructor(s), administrator(s) as shown in Fig. 1. Based on the threat outcome on the critical system components, recovery cost and participants, we have categorized the impact level

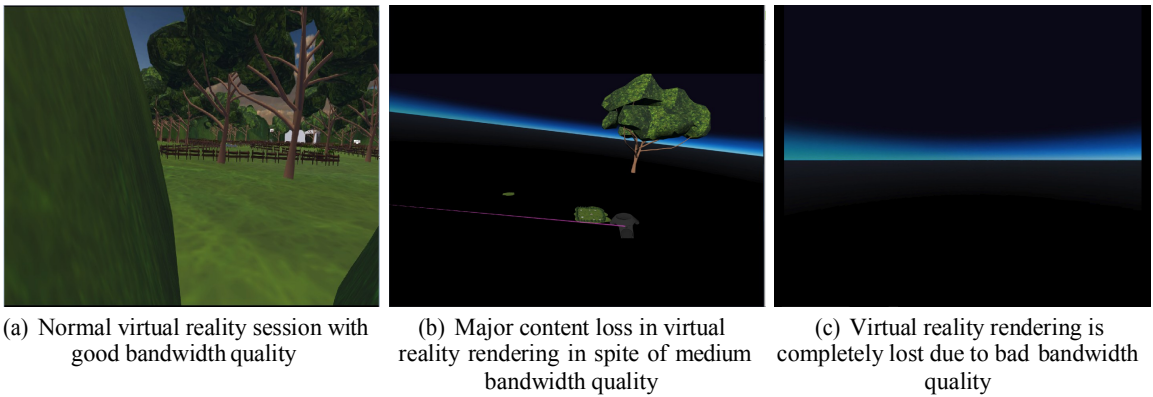

(a) Normal virtual reality session with good bandwidth quality

(b) Major content loss in virtual reality rendering in spite of medium bandwidth quality

(c) Virtual reality rendering is completely lost due to bad bandwidth quality

Fig. 10: Effect of high, medium and low bandwidth quality on virtual reality rendering.

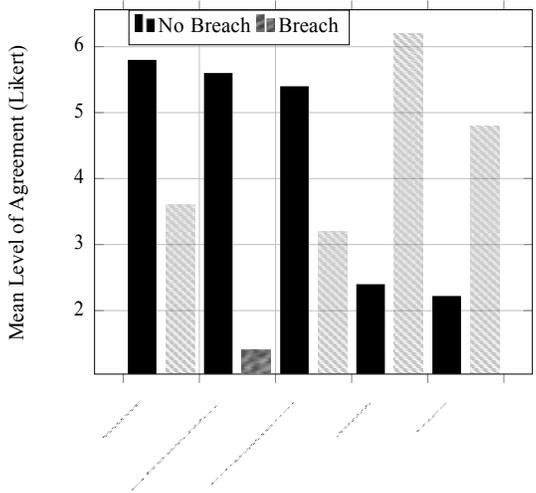

Fig. 11: Differences in experiences for Breach and No Breach scenarios.

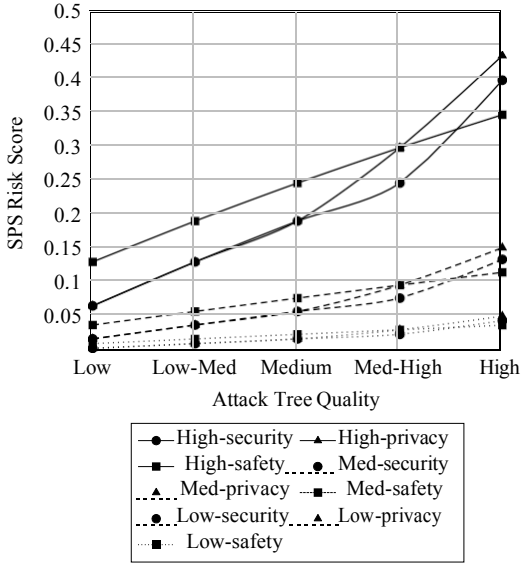

Fig. 13: Risk Score characteristics based on the variation in quality of attack trees for all the SPS trees.

TABLE IV: Impact scale of threats on vSocial security.

| Attack Events | Scale of Impact |
| --- | --- |
| XSS attack | 4 |
| Delay Packets | 2 |
| DoS | 5 |
| DDoS | 5 |
| Packet Loss | 4 |
| Instructor Spoofing | 2-3 |
| Admin Spoofing | 4-5 |
| Get Password/Unauthorized Login | 2-3 |
| SQL injection | 3 |
| Spoofing | 2 |

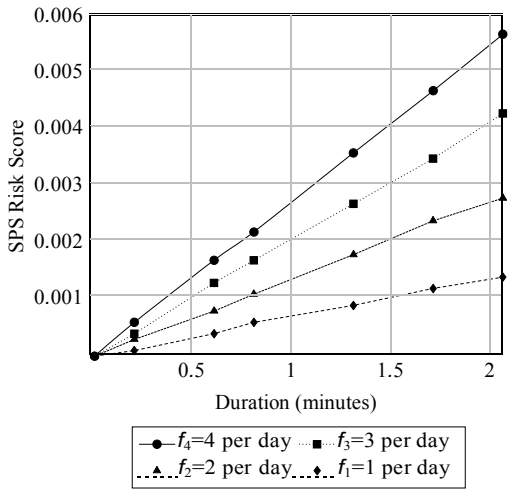

Fig. 12: Characteristic graphs showing Security Risk Score w.r.t. threat behavior including duration (minutes) and rate of threat occurrence.

with the scale ranging from 1 (lowest) to 5 (highest). The least impact threats with score of 1-2 can be easily repaired, whereas threats with significant impact with score of 4-5 require more time and cost. We assigned the impact level scores to each threat for safety in Table III, security in Table IV and privacy in Table V.

Selected results from Table III are shown as concrete examples. Redirecting packets to the malicious server can lead to session takeover or a change in the student and instructor

TABLE V: Impact scale of threats on vSocial privacy.

| Attack Events | Scale of Impact |
|---|---|
| SQL Injection | 3 |
| Undeleted Account | 1 |
| Change Packet Destination | 4 |
| Packet Sniffing | 5 |
| Screen Observation | 1 - 2 |
| Hand Movement Observation | 1 - 2 |
| User not notified | 4 |

VR view(s). Therefore, high impact level of 4 is assigned to this attack in Table III. On the other hand, minor programming errors or bugs due to poorly written code can cause only minor issues, and hence result in a impact level of 1. Medium impact threats are those that are not as severe as redirecting packets, but not as mild as poorly written code. Example is the occurrence of extended VR sessions, which could result in mental fatigue or cybersickness of a student.

V. CONCLUSION AND FUTURE WORK

In this paper, we propose a risk assessment framework to address the SPS challenges in a VRLE application system. We employ formalized attack trees to map threats and associated risks to the various system modules to generate a corresponding risk score. To show the effectiveness of our framework, we introduced three sample attack experiments (network discrepancy, packet loss and sniffing) on a realistic vSocial testbed. We modified the quality of attack trees (well-designed vs. adhoc) based on the number and extent of threats being considered. A trade-off analyses between the adhoc and well-designed attack trees was performed. We found that - although the cost of managing the risk analysis increases, our approach provides realistic insights for risks associated with the system component vulnerabilities to inform VRLE policy management by an administrator to mitigate risks.

As part of future work, our risk assessment framework can be extended to investigate policy change control during VRLE sessions for ensuring a secure VR application that protects user privacy and ensures user safety.

REFERENCES

[1] C. Zizza, A. Starr, D. Hudson, S. S. Nuguri, P. Calyam and Z. He, "Towards a Social Virtual Reality Learning Environment in High Fidelity", *IEEE Annual Consumer Communications & Networking Conference (CCNC)*, 2018.
[2] J. Jia, W. Chen, "The Ethical Dilemmas of Virtual Reality Application in Entertainment", *IEEE Intl. Conference on Computational Science and Engineering (CSE)*, 2017.
[3] J. P. Stichter, M. J. Herzog, K. Visovsky, C. J. Schmidt, T. Randolph, T. Schultz, N. Gage, "Social Competence Intervention for Youth with Asperger Syndrome and High-functioning Autism: An Initial Investigation", *Journal of Autism and Developmental Disorders*, Vol. 40, No. 9, pp. 1067-1079, 2010.
[4] B. Fineman, N. Lewis, "Securing Your Reality: Addressing Security and Privacy in Virtual and Augmented Reality Applications", *EDUCAUSE Review*. [Online]. Available: https://er.educause.edu/articles/2018/5/securing-your-reality-addressing-security-and-privacy-in-virtual-and-augmented-reality-applications [Accessed July 30, 2018].
[5] Y. Wei, X. Wang, P. Calyam, D. Xuan, W. Zhao, "On detecting camouflaging worm", *IEEE ASAC*, 2006.
[6] W. Zhou, Y. Jia, A. Peng, Y. Zhang, P. Liu, "The Effect of IoT New Features on Security and Privacy: New Threats, Existing Solutions, and Challenges Yet to Be Solved", *IEEE Internet of Things Journal*, 2018.
[7] K. Fu, T. Kohno, D. Lopresti, E. Mynatt, K. Nahrstedt, S. Patel, D. Richardson, B. Zorn, "Safety, Security, and Privacy Threats Posed by Accelerating Trends in the Internet of Things", *Computing Community Consortium (CCC) Technical Report*, Vol. 29, No. 3, 2017.
[8] D. Franziska Roesner, T. Kohno, D. Molnar, "Security and Privacy For Augmented Reality Systems", *Communications of the ACM*, Vol. 57, No. 4, pp. 88-96, 2014.
[9] R. Roman, J. Zhou, J. Lopez,"On the Features and Challenges of Security and Privacy in Distributed Internet of Things", *Elsevier Computer Networks Journal*, Vol. 57, No. 10, pp. 2266-2279, 2013.
[10] Y. Yang, L. Wu, G. Yin, L. Li, H. Zhao, "A Survey on Security and Privacy Issues in Internet-of-Things", *IEEE Internet of Things Journal*, Vol. 4, No. 5, pp. 1250-1258, 2017.
[11] R. McPherson, S. Jana, V. Shmatikov "No Escape from Reality: Security and Privacy of Augmented Reality Browsers", *Proc. of the 24th Intl. Conference on World Wide Web*, 2015.
[12] S. Yi, Z. Qin, Q. Li, "Security and Privacy Issues of Fog Computing: A Survey", *Proc. of Intl. Conference on Wireless Algorithms, Systems, and Applications*, 2015.
[13] M. Haris, H. Haddadi, P. Hui "Privacy Leakage in Mobile Computing Tools, Methods, Characteristics", *arXiv preprint arXiv:1410.4978*, 2014.
[14] K. Stanney, R. Murant, R. Kennedy, "Human Factor Issues in Virtual Environments: A Review of the Literature", *Presence: Teleoperators and Virtual Environments*, Vol. 7, No. 4, pp. 327-351, 1998.
[15] M. Dennison, A. Wisti, M. D'Zmura, "Use of Physiological Signals to Predict Cybersickness", *Displays*, Vol. 44, pp. 42-52, 2016.
[16] N. Glaser, M. Scmidt, "Usage Considerations of 3D Collaborative Virtual Learning Environments to Promote Development and Transfer of Knowledge and Skills for Individuals with Autism", *Technology, Knowledge and Learning*, 2018.
[17] P. Ralston, J. Graham, J. Hieb, "Cyber Security Risk Assessment for SCADA and DCS Networks", *ISA Transactions*, Vol. 46, No. 4, pp. 583-594, 2007.
[18] S. Mauw, M. Oostdijk, "Foundations of Attack Trees", *Intl. Conference on Information Security and Cryptology*, pp. 1-14, 2018.
[19] K. Edge, R. Raines, M. Grimaila, R. Baldwin, R. Bennington, C. Reuter, "The Use of Attack and Protection Trees to Analyze Security for an Online Banking System", *IEEE Annual Hawaii Intl. Conference*, 2007.
[20] H. Kong, M. Hong, T. Kim, "Security Risk Assessment Framework for Smart Car using the Attack Tree Analysis", *Intl. Conference on Innovative Mobile and Internet Services in Ubiquitous Computing (IMIS)*, 2016.
[21] M. Fraile, M. Ford, O. Gadyatskaya, R. Kumar, M. Stoelinga, R. Trujillo-Rasua, "Using Attack-Defense Trees to Analyze Threats and Countermeasures in an ATM: A Case Study", *IFIP Working Conference on The Practice of Enterprise Modeling*, 2016.
[22] A. Marback, H. Do, K. He, S. Kondamarri, D. Xu, "A Threat Model-based Approach to Security Testing", *Software: Practice and Experience*, Vol. 43, No. 2, pp. 241-258, 2012.
[23] A. Gholami, E. Laure, "Advanced Cloud Privacy Threat Modeling", *Procedia Computer Science*, Vol. 37, pp. 489-496, 2014.
[24] A. Almulhem, "Threat Modeling for Electronic Health Record Systems", *Journal of Medical Systems*, Vol. 36, No. 5, pp. 2921-2926, 2014.
[25] R. Hasan, S. Myagmar, A. Lee, W. Yurcik, "Toward a Threat Model for Storage Systems", *Proc. of the ACM Workshop on Storage Security and Survivability*, pp. 94-102, 2005.
[26] "SecurITree for Attack Tree analysis". [Online] Available at: https://www.amenaza.com [Accessed July 30 2018].
[27] M. Berman, J. Chase, L. Landweber, A. Nakao, M. Ott, D. Raychaudhuri, R. Ricci, I. Seskar, "GENI: A Federated Testbed for Innovative Network Experiments", *Computer Networks*, Vol. 61, pp. 5-23, 2014.
[28] "High Fidelity", 2018. [Online]. Available: https://highfidelity.com. [Accessed July 30 2018].
[29] "SteamVR", 2018. [Online]. Available: http://store.steampowered.com/steamvr. [Accessed July 30 2018].
[30] C. Tao,"Clumsy 0.2", 2018. [Online]. Available: https://jagt.github.io/clumsy. [Accessed July 30 2018].
[31] P. Calyam, M. Haffner, E. Ekici, C. G. Lee, "Measuring Interaction QoE in Internet Videoconferencing", *IEEE/IFIP Management of Multimedia and Mobile Networks and Services (MMNS)*, 2007.